# Multiphasic profiles for the biologically important ion activities of NaCl and KCl

Per Nissen

Norwegian University of Life Sciences
Department of Ecology and Natural Resource Management
P. O. Box 5003, NO-1432 Ås, Norway

per.nissen@nmbu.no

# Introduction

When plotted in linear transformations of the Michaelis-Menten equation, ion uptake in plants has been shown to be multiphasic, i.e. to be represented by a series of straight lines separated by discontinuous transitions (Nissen 1971, 1974, 1991, 1996). Reanalysis of data for other transport, binding and enzyme systems has also given such kinetics (Nissen and Martín-Nieto 1998). Recently, such profiles have been found for a variety of biological as well as non-biological processes and phenomena, including activation of ion channels, binding, pH, folding/unfolding, effects of various interactions and chain length (Nissen 2015a,b, Nissen 2016a-d). These biological and nonbiological processes have little in common beyond the ions involved. Here I show that the multiphasic property seen in so many systems is present in the fundamental physical properties of the ion themselves.

The characteristics of multiphasic profiles have now also been found in profiles for activities of ions in simple inorganic solutions, indicating that they are somehow causing the multiphasic profiles in more complex systems. A set of extensive and precise data (Lee et al. 2002) for the activities of the biologically very important ions $Na^+$, $K^+$ and $Cl^-$ (in NaCl and KCl) at four different temperatures (15, 25, 35 and 45$^o$C) will be reanalyzed in the present paper. The activities of NaBr and KBr will be plotted in a forthcoming paper. Extensive data for these and other ion activities Wilczek-Vera et al. (2004) can also be precisely represented by multiphasic profiles (in preparation).

# Reanalysis

Data for the activity coefficients of $Na^+$ and $Cl^-$ in aqueous solutions of various concentrations at 288.15, 298.15, 308.15 and 318.15 K (Tables 1 and 6 in Lee et al. 2002) have been plotted against each other (Figs 1-8). Data for the activity coefficients of $K^+$ and $Cl^-$ (Tables 2 and 5 in Lee et al. 2002) have also been plotted (Figs 9-16).

As shown by the very high absolute r values for the straight lines (47 of the 58 r values in Figs 1-16 are 0.999 or higher), the data can, without exception, be well represented by straight lines. When lines intersect in a common point, the transition is obviously discontinuous (intersection between lines VII and VIII in Fig. 3, between lines I and II and between lines II and III in Fig. 5, between lines V and VI in Fig. 6, between lines I and II in Fig. 7, between lines III and IV in Fig. 9). When lines intersect close to a point, the transition is probably also discontinuous.

Adjacent lines are quite often parallel or nearly so and are then necessarily noncontiguous, i.e. the transition is in the form of a jump (lines IV and V in Fig. 3, lines V and VI in Fig. 4, lines III, IV and VI in Fig. 5, lines IV and V in Fig. 6, lines III and IV and lines V and VI in Fig. 8, lines II and III and lines V and VI in Fig. 10, lines III and IV and lines V and VI in Fig. 11, lines IV and V in Fig. 13, lines V and VI in Fig. 14, lines V and VI in Fig. 15, lines IV and V in Fig. 16). Many of the jumps are tiny, but the often high r values of the adjacent lines indicate that the assignments are correct.

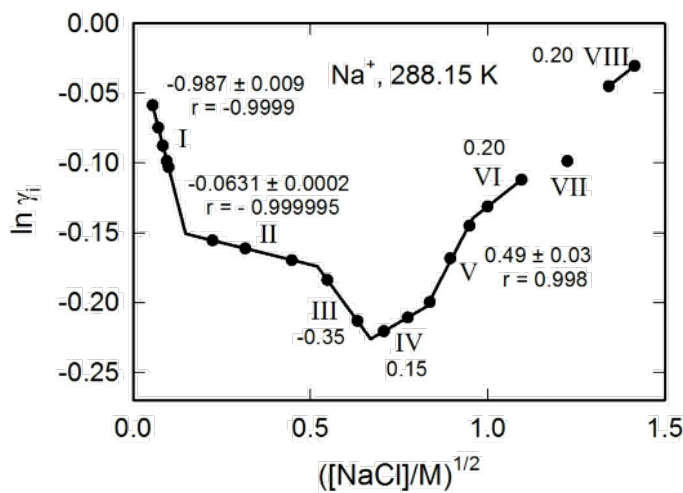

**Fig. 1.** Eight phases. Transitions at 0.148, 0.519, 0.670, 0.828 and 0.957, between 1.095 and 1.225, and between 1.225 and 1.342. Insufficiently detailed data for resolution of phase VII. Lines IV, VI and VIII are parallel, precisely so for VI and VIII. Very and exceedingly high absolute r values for lines I and II.

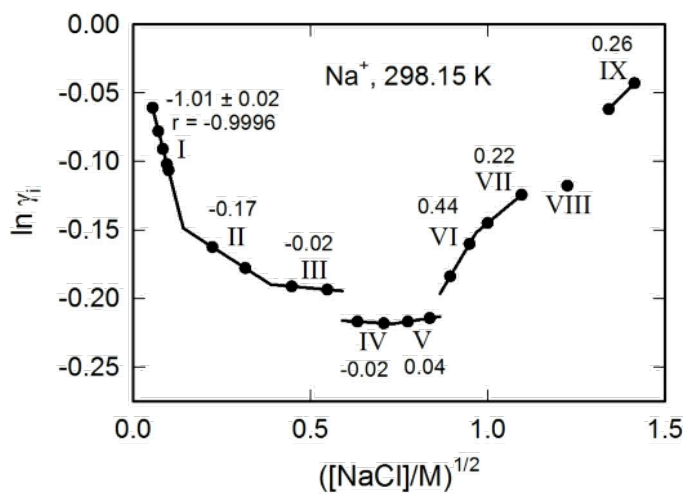

**Fig. 2.** Nine phases. Transitions at 0.143 and 0.388, between 0.548 and 0.632 (jump), at 0.732, between 0.837 and 0.894 (jump), at 0.967, between 1.095 and 1.225, and between 1.225 and 1.342. Insufficiently detailed data for resolution of phase VIII. Lines III and IV are parallel, as are approximately also lines VII and IX. Very high absolute r value for line I. Lines IV and V are probably not a single line (r = 0.685.

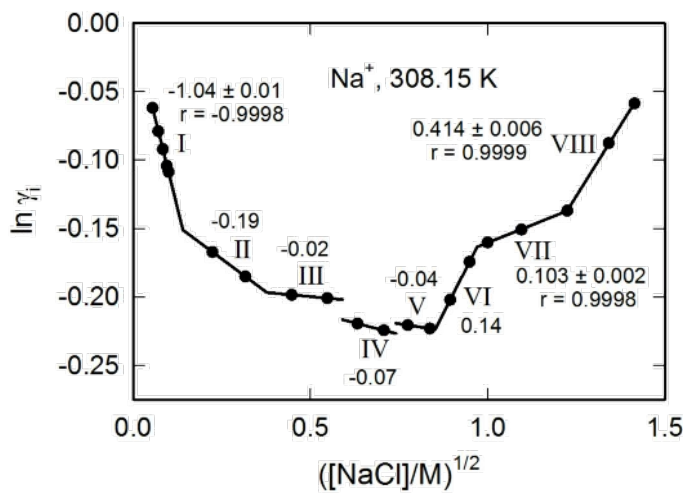

**Fig. 3.** Eight phases. Transitions at 0.140 and 0.377, between 0.548 and 0.632 (jump), between 0.707 and 0.775 (jump), and at 0.852, 0.970 and 1.225. Lines III-V are roughly parallel. Very high absolute r values for lines I, VII and VIII.

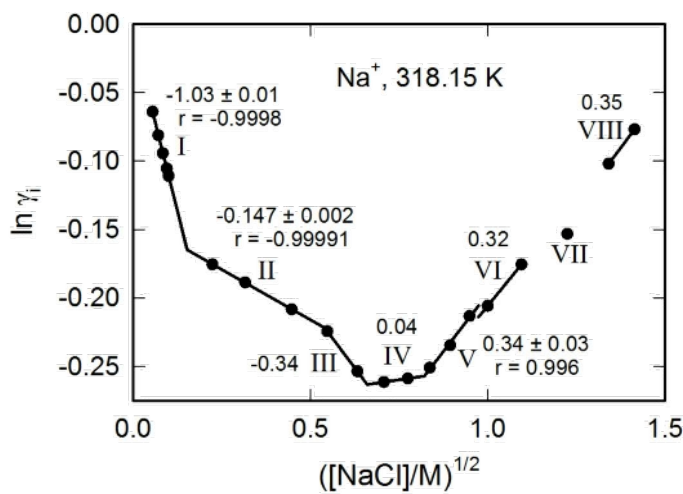

**Fig. 4.** Eight phases. Transitions at 0.152, 0.540, 0.660 and 0.822, between 0.949 and 1.000 (small jump), between 1.095 and 1.225, and between 1.225 and 1.342. Insufficiently detailed data for resolution of phase VII. Lines V, VI and VIII are parallel or approximately so. Very high absolute r values for lines I and II.

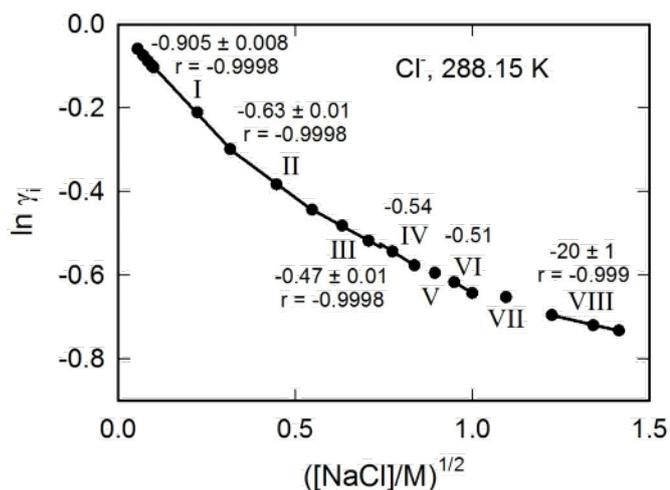

**Fig. 5.** Eight phases. Transitions at 0.316 and 0.548, between 0.707 and 0.775 (small jump), between 0.837 and 0.894, between 0.894 and 0.949, between 1.000 and 1.095, and between 1.095 and 1.225. Insufficiently detailed data for resolution of phases V and VII. Lines III, IV and VI are about parallel. Very high absolute r values for lines I-III.

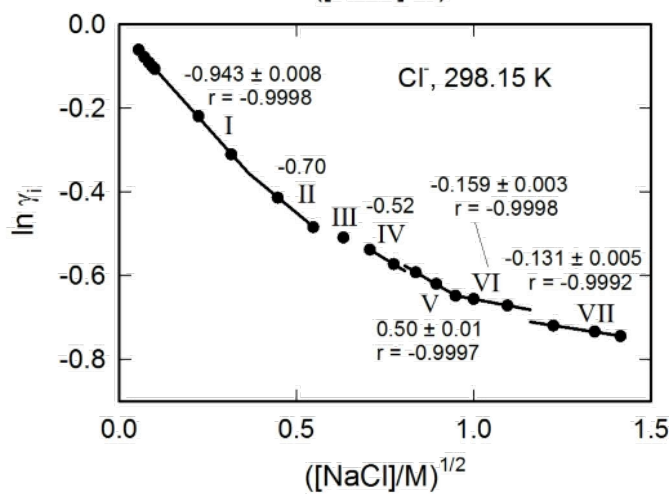

**Fig. 6.** Seven phases. Transitions at 0.369, between 0.548 and 0.632, between 0.632 and 0.707, between 9.775 and 0.837 (tiny jump), at 0.949, and between 1.095 and 1.225 (jump). Insufficiently detailed data for resolution of phase III. Lines IV and V are parallel. Very high absolute r values for Lines I, V, VI and VII.

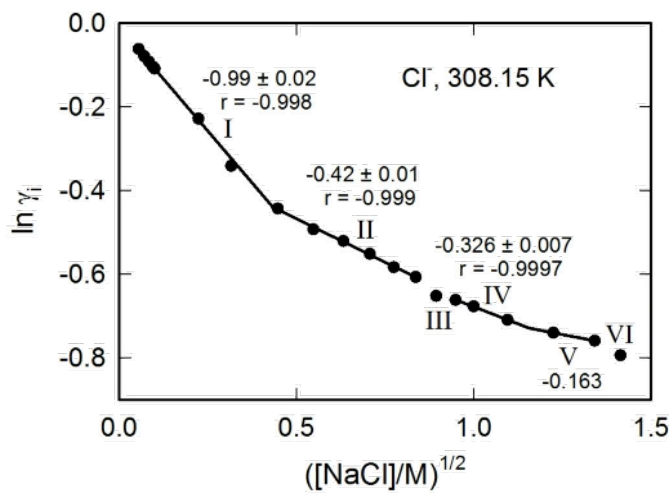

**Fig. 7.** Six phases. Transitions at 0.447, between 0.837 and 0.894, between 0.894 and 0.949, at 1.157, and between 1.342 and 1.414. Insufficiently detailed data for resolution of phases III and VI. High to very high absolute r values for lines I, II and IV. There may possibly be two phases in the range of line I (r for the first 6 points = -0.9998).

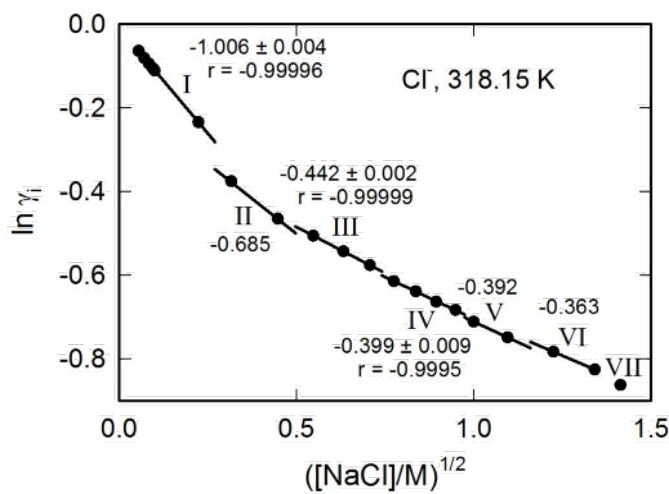

**Fig. 8.** Seven phases. Transitions between 0.224 and 0.316 (jump), between 0.447 and 0.548 (tiny jump), between 0.707 and 0.775 (tiny jump), between 0.949 and 1.000 (tiny jump), between 1.095 and 1.225 (tiny jump), and between 1.342 and 1.414. Insufficiently detailed data for resolution of phase VII. Lines IV and V are parallel. Very to exceedingly high absolute r values for lines I, III and IV.

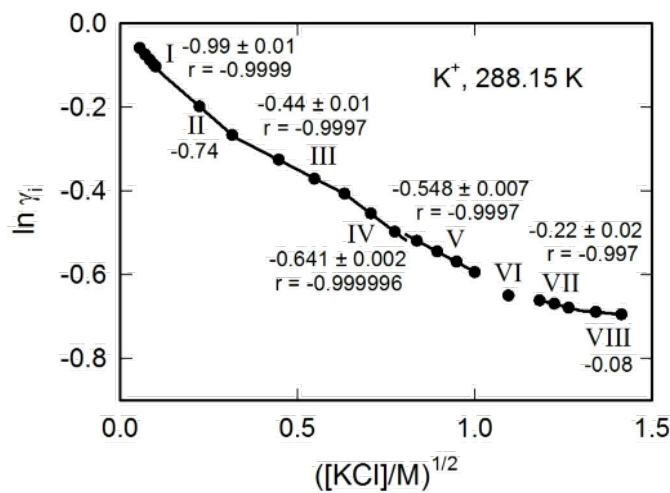

**Fig. 9.** Eight phases. Transitions at 0.116, 0.324 and 0.632, between 0.775 and 0.837 (tiny jump), between 1.000 and 1.095, between 1.095 and 1.183, and at 1.303. Insufficiently detailed data for resolution of phase VI. Very high absolute r values for lines I, III, IV and V. A single line in the range of phases VI-VIII seems unlikely (r = -0.990).

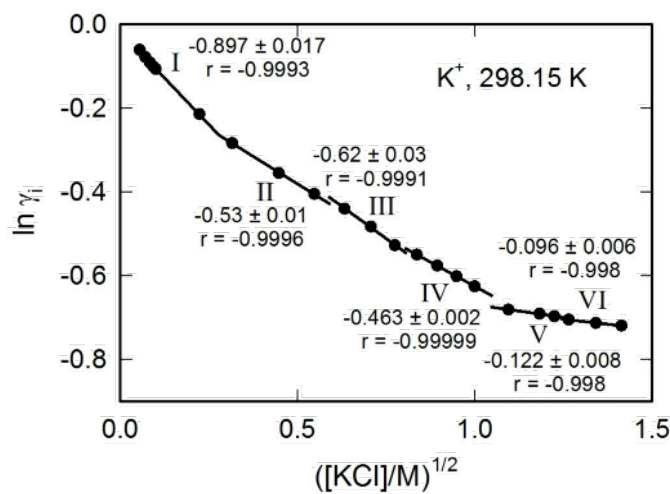

**Fig. 10.** Six phases. Transitions at 0.278, between 0.548 and 0.632 (tiny jump), between 0.775 and 0.837 (tiny jump), between 1.000 and 1.095 (tiny jump), and between 1.22 and 1265 (jump, not visible). Lines II-IV are roughly parallel, but the very high absolute r values show that the lines are separate.

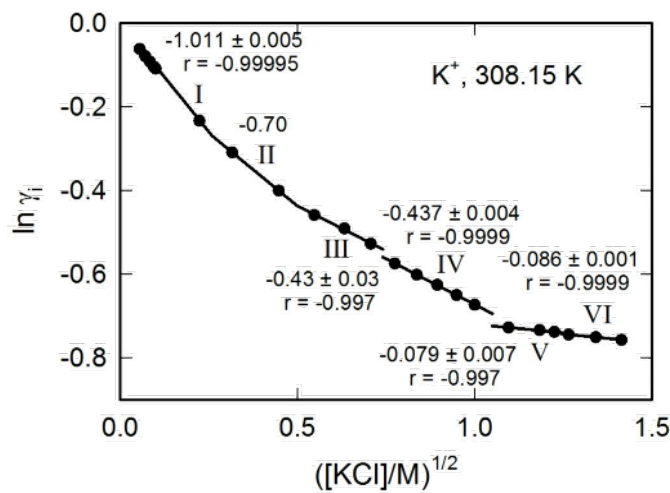

**Fig. 11.** Six phases. Transitions at 0.259 and 0.499, between 0.707 and 0.775 (small jump), between 1.000 and 1.095 (jump), and between 1.225 and 1.265 (jump, not visible). Lines III and IV are precisely parallel, as are lines V and VI. Very high absolute r values for lines I, IV and VI.

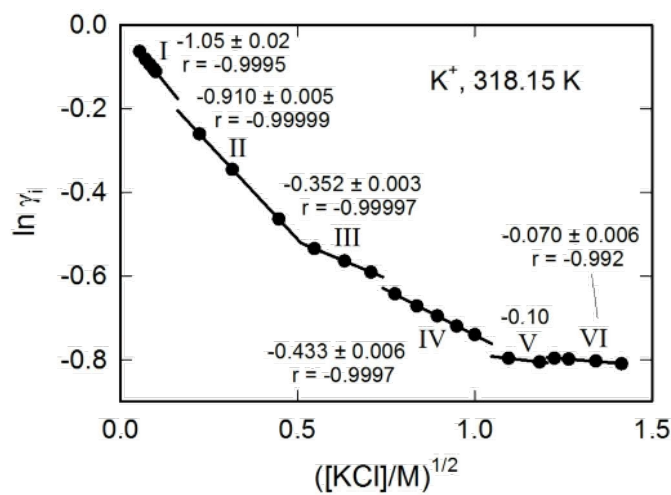

**Fig. 12.** Six phases. Transitions between 0.100 and 0.224 (jump), at 0.510, between 0.707 and 0.775 (jump), between 1.000 and 1.095, and between 1.183 and 1.225 (tiny jump). Lines I and II are about parallel, as are lines V and VI. Very high absolute r values for lines I-IV.

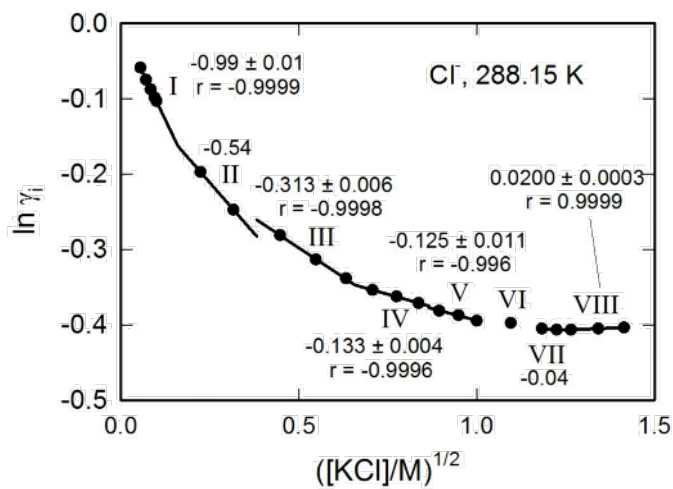

**Fig. 13.** Eight phases. Transitions at 0.160, between 0.316 and 0.447 (jump), at 0.658, between 0.837 and 0.894 (tiny jump), between 1.000 and 1.095, between 1.095 and 1.183, and at 1.239. Insufficiently detailed data for resolution of phase VI. Lines IV and V are parallel, but probably not a single line (r = -0.999 vs. r = -0.9996 for line IV.) Very high absolute r values for lines I, III, IV and VIII.

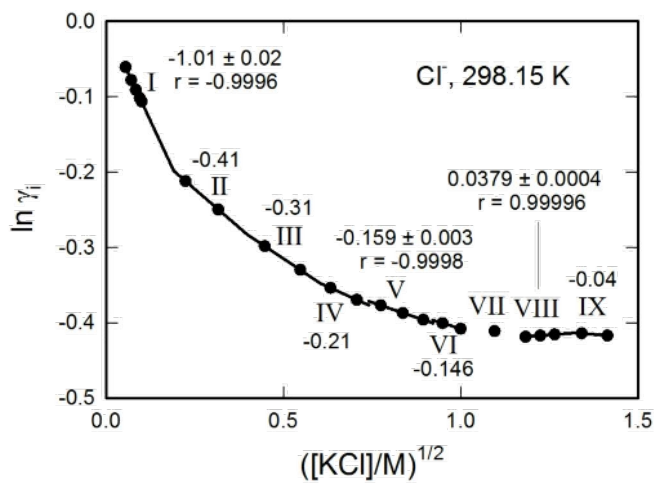

**Fig. 14.** Nine phases. Transitions at 0.190, 0.398 and 0.606, between 0.707 and 0.775 (tiny jump), between 0.894 and 0.949 (tiny jump), between 1.000 and 1.095, between 1.095 and 1.183, and at 1.324. Insufficiently detailed data for resolution of phase VII. Lines IV-VI are parallel or approximately so, but are probably not a single line (r = -0.995). Very high absolute r values for lines I, V and VIII.

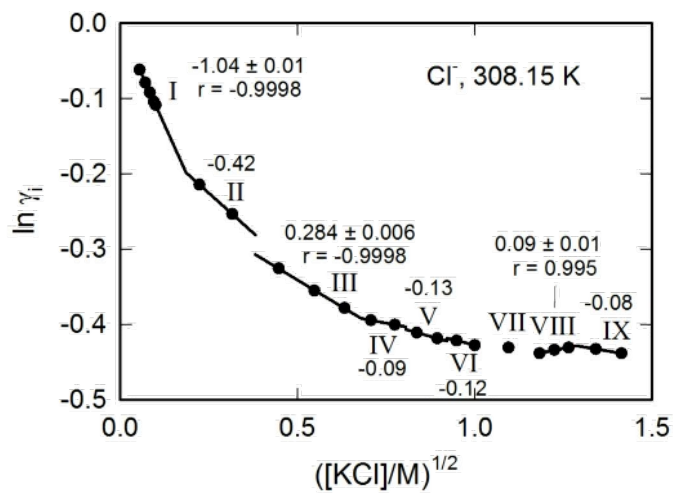

**Fig. 15.** Nine phases. Transitions at 0.186, between 0.316 and 0.447, between 0.775 and 0.837 (tiny jump), between 0.894 and 0.949 (tiny jump), between 1.000 and 1.095, between 1.095 and 1.183, and at 1.287. Insufficiently detailed data for resolution of phase VII. Lines V and VI are parallel, as is approximately also line IV. Lines IV-VI are probably not a single line (r = -0.992). Very high absolute r values for lines I and III.

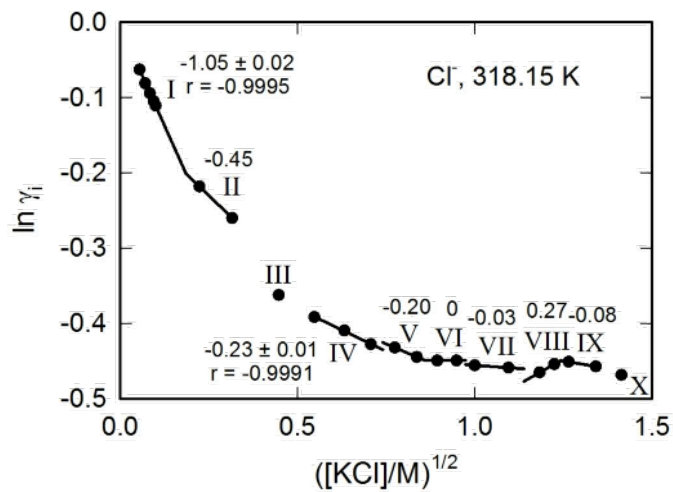

**Fig. 16.** Ten phases. Transitions at 0.185, between 0.316 and 0.447, between 0.447 and 0.548, between 0.707 and 0.775 (tiny jump), at 0.860, between 0.949 and 1.000 (tiny jump), between 1.095 and 1.183 (jump), at 1.243, and between 1.342 and 1.414. Insufficiently detailed data for resolution of phases III and X. Lines IV and V are about parallel, as are lines VI and VII (line VI is horizontal). Lines VI and VII are probably not a single line (r = -0.944). Very high absolute r values for lines I and IV.

As concluded by the authors, the ion activity coefficients decreased with increasing temperature. However, in contrast to their representation (Figs 2 and 3 in Lee et al. 2002) of the profiles for ion activities as curvilinear, the profiles are precisely multiphasic. The profiles for $Na^+$ (from NaCl) can be represented by 8 or 9 phases (Figs 1-4). The profiles has a minimum at 0.632-0.837 and increase again at higher concentrations. The profiles for $Cl^-$ (from NaCl) can be represented by 6-8 phases (Figs 5-8) and decrease with increasing ion concentration. The profiles for $K^+$ (from KCl) can be represented by 6 phases for the three highest temperatures (Figs 10-12), but by 8 phases for the lowest temperature (Fig. 9). In marked contrast to the profiles for $Na^+$, the ion activities decreased over the entire concentration range. The profiles for $Cl^-$ (from KCl) can be represented by 8-10 phases (Figs 13-16). Again, the activities decreased with increasing concentrations, maybe except for a slight increase at a few of the highest concentrations. The profiles for any one ion are very similar at the various temperatures, and it seems that any differences in the patterns may be accidental.

## Conclusions and Questions

As shown, the present data are very if not exceedingly precise. It seems that they allow some far-reaching conclusions, but they do also present difficult questions. In many experimental studies, in non-biological as well as in biological systems, the systems have been taken to be continuous as a function of the parameter being varied, without any disruptions. However, it has now been shown (see Introduction) that many systems are in fact discontinuous rather than continuous. There has been little or no awareness of these discontinuities which seem to originate from discontinuities in ion activities in the system being studied. Clearly, such studies should now be carried out with sufficient detail and precision for any discontinuities to be recognized.

The molecular basis of the discontinuities remains unclear. Why should there be clear and reproducible discontinuities in simple salt solutions? Why should the lines be straight, apparently perfectly so? Why are adjacent lines quite often parallel or nearly so? Why do the profiles for $Na^+$ and $K^+$ differ so markedly? To what extent does the finding of multiphasic profiles invalidate conclusions from less thorough studies?

**Acknowledgment** – I am very grateful to Bob Eisenberg for his continued interest and encouragement.